\documentclass[sigconf]{acmart}

\AtBeginDocument{%
  \providecommand\BibTeX{{%
    \normalfont B\kern-0.5em{\scshape i\kern-0.25em b}\kern-0.8em\TeX}}}

\acmConference[ICSE 2024]{46th International Conference on Software Engineering}{April 2024}{Lisbon, Portugal}

\acmBooktitle{} 
\acmPrice{}
\acmISBN{}
\usepackage[linesnumbered,boxed,ruled,commentsnumbered]{algorithm2e}
\usepackage{multirow}
\usepackage{multicol}
\usepackage{rotating}
\usepackage{xspace}
\usepackage{xcolor}
\usepackage{colortbl}
\usepackage{color}
\usepackage{enumitem}
\usepackage{listings}
\usepackage{threeparttable}
\usepackage{graphicx}
\usepackage{subfigure}
\SetKwInput{KwInput}{Input}                
\SetKwInput{KwOutput}{Output}              
\SetKwFunction{FRecurs}{FnRecursive}%

\newcolumntype{L}[1]{>{\raggedright\let\newline\\\arraybackslash\hspace{0pt}}m{#1}}
\newcolumntype{C}[1]{>{\centering\let\newline\\\arraybackslash\hspace{0pt}}m{#1}}
\newcolumntype{R}[1]{>{\raggedleft\let\newline\\\arraybackslash\hspace{0pt}}m{#1}}

\definecolor{codegreen}{rgb}{0,0.6,0}
\definecolor{codegray}{rgb}{0.5,0.5,0.5}
\definecolor{codepurple}{rgb}{0.58,0,0.82}
\definecolor{backcolour}{rgb}{0.95,0.95,0.92}

\lstdefinestyle{mystyle}{
    commentstyle=\color{codegreen},
    keywordstyle=\color{magenta},
    numberstyle=\tiny\color{codegray},
    stringstyle=\color{codepurple},
    basicstyle=\footnotesize,
    breakatwhitespace=false,
    breaklines=true,
    captionpos=b,
    keepspaces=true,
    showspaces=false,
    showstringspaces=false,
    showtabs=false,
    tabsize=2
}

\setlist{noitemsep} 

\definecolor{darkpastelred}{rgb}{0.76, 0.23, 0.13}
\definecolor{ao(english)}{rgb}{0.0, 0.5, 0.0}

\lstdefinelanguage{diff}{
  morecomment=[f][\color{blue}]{@@},     
  morecomment=[f][\color{red}]-,         
  morecomment=[f][\color{codegreen}]+,       
  morecomment=[f][\color{red}]{---}, 
  morecomment=[f][\color{codegreen}]{+++},
}

\newcommand{\toolname}{{\bf P-EPR}\xspace}

\copyrightyear{2024}
\acmYear{2024}
\setcopyright{acmlicensed}\acmConference[ICSE '24]{2024 IEEE/ACM 46th International Conference on Software Engineering}{April 14--20, 2024}{Lisbon, Portugal}
\acmBooktitle{2024 IEEE/ACM 46th International Conference on Software Engineering (ICSE '24), April 14--20, 2024, Lisbon, Portugal}
\acmPrice{15.00}
\acmDOI{10.1145/3597503.3623310}
\acmISBN{979-8-4007-0217-4/24/04}

\newboolean{showcomments}
\setboolean{showcomments}{true}
\ifthenelse{\boolean{showcomments}}
 { \newcommand{\mynote}[2]{
      \fbox{\bfseries\sffamily\scriptsize#1}
        {\small$\blacktriangleright$\textsf{\emph{#2}}$\blacktriangleleft$}}}
        { \newcommand{\mynote}[2]{}}

\newcommand{\eat}[1]{}

\begin{document}

\title{Practical Program Repair \textit{via} Preference-based Ensemble Strategy}

\author{Wenkang Zhong}
\affiliation{
  \institution{National Key Laboratory for Novel Software Technology, Nanjing University, China}
  \city{Nanjing}
  \country{China}}
\email{zhongwenkang97@foxmail.com}

\author{Chuanyi Li}
\authornote{Corresponding authors.}
\affiliation{
  \institution{National Key Laboratory for Novel Software Technology, Nanjing University, China}
  \city{Nanjing}
  \country{China}}
\email{lcy@nju.edu.cn}

\author{Kui Liu}
\affiliation{
  \institution{Huawei Software Engineering Application Technology Lab}
  \city{Hangzhou}
  \country{China}}
\email{brucekuiliu@gmail.com}

\author{Tongtong Xu}
\affiliation{
  \institution{Huawei Software Engineering Application Technology Lab}
  \city{Hangzhou}
  \country{China}}
\email{xutongtong9@huawei.com}

\author{Jidong Ge}
\authornotemark[1]
\affiliation{
  \institution{National Key Laboratory for Novel Software Technology, Nanjing University, China}
  \city{Nanjing}
  \country{China}}
\email{gjd@nju.edu.cn}

\author{Tegawendé F. Bissyandé}
\affiliation{
  \institution{University of Luxembourg}
  \country{Luxembourg}}
\email{tegawende.bissyande@uni.lu}

\author{Bin Luo}
\affiliation{
  \institution{National Key Laboratory for Novel Software Technology, Nanjing University, China}
  \city{Nanjing}
  \country{China}}
\email{luobin@nju.edu.cn}

\author{Vincent Ng}
\affiliation{
  \institution{Human Language Technology Research Institute, University of Texas at Dallas}
  \city{Richardson}
  \state{Texas}
  \country{USA}}
\email{vince@hlt.utdallas.edu}

\begin{abstract}

To date, over 40 Automated Program Repair (APR) tools have been designed with varying bug-fixing strategies, which have been demonstrated to have complementary performance in terms of being effective for different bug classes. Intuitively, it should be feasible to improve the overall bug-fixing performance of APR via assembling existing tools.
Unfortunately, simply invoking all available APR tools for a given bug can result in unacceptable costs on APR execution as well as on patch validation (via expensive testing). Therefore, while assembling existing tools is appealing, it requires an efficient strategy to reconcile the need to fix more bugs and the requirements for practicality.
In light of this problem, we propose a \underline{\textbf{P}}reference\underline{\textbf{-}}based \underline{\textbf{E}}nsemble \underline{\textbf{P}}rogram \underline{\textbf{R}}epair framework (\toolname), which seeks to effectively rank APR tools for repairing different bugs. {\bf P-EPR} is the first non-learning-based APR ensemble method that is novel in its exploitation of repair patterns as a major source of knowledge for ranking APR tools and its reliance on a dynamic update strategy that enables it to immediately exploit and benefit from newly derived repair results. Experimental results show that \toolname outperforms existing strategies significantly both in flexibility and effectiveness.

\end{abstract}



 \maketitle

\section{Introduction}
\label{sec:introduction}

Bug fixing is a challenging, time-consuming, and labor-intensive task, often consuming a significant portion of developers' efforts~\cite{bugfix_consume}. 
To address this challenge, Automated Program Repair (APR) \cite{suvey18_APR} has been dedicated to automatically fixing bugs without human intervention, and has become a hot field in the software engineering community. To date, more than 40 APR tools have been proposed as the momentum for program repair is growing.

\begin{figure}[t]
\setlength{\abovecaptionskip}{0cm}
  \centering
  \setlength{\abovecaptionskip}{0cm}
  \includegraphics[width=0.9\linewidth]{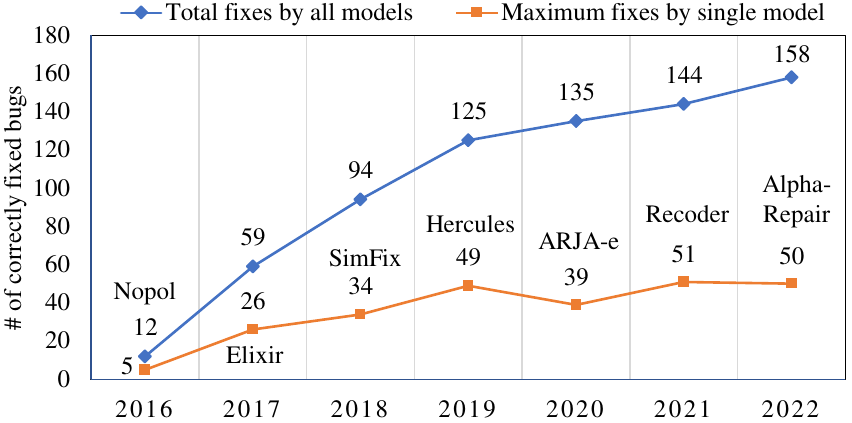}
  \caption{Fixing capabilities of single top-performing APR system vs. their integration over the years (2016-2022) on Defects4J v1.2 bugs under normal fault localization setting.
  } 
  \label{fig:performanceReview}
  \vspace{-2mm}
\end{figure}

Practitioners have been exploring advanced techniques that could overwhelmingly outperform all the other APR techniques in all lines of bug-fixing performance.
Nevertheless, various experimental results in the literature suggest that there is 
at least one APR tool whose bug-fixing merit cannot be achieved by other APR tools~\cite{FSE19_empAPR,icse20_efficiency,liu2021critical,ese21_EAPR}.
Different APR tools present complementary repairability to each other.
For example, as the recent state-of-the-art APR tool, AlphaRepair~\cite{fse22_alpharepair} can correctly fix 50 Defects4J bugs under normal fault localization setting, while there are 108 Defects4J bugs that cannot be fixed by it but can be correctly fixed by other APR tools, as shown in Figure \ref{fig:performanceReview}. 

Given the results in Figure \ref{fig:performanceReview}, it is tempting to try all APR tools so that more bugs can be fixed, but trying all APR tools to fix a given bug is simply impractical
due to the unacceptable costs of tool invocation and patch validation. 
For instance, Durieux et al. \cite{FSE19_empAPR} spent 314 days executing and validating 11 APR tools even with Grid'5000 \cite{Grid'5000}.
Additionally, a single APR tool could generate some plausible patches for a given bug~\cite{qi2015analysis}, and more APR tools would generate more plausible patches~\cite{icse20_efficiency}, which will considerably increase the difficulty of distinguishing the correct patch from plausible ones. For this problem, the current state-of-the-art strategy, E-APR \cite{ese21_EAPR}, frames the selection as a supervised machine learning task, specifically a multi-label classification task, that involves identifying the set of tools that should be used to fix a given bug from an ensemble of tools.
However, training E-APR requires a large amount of labeled data, which is costly to obtain in the program repair field (manually validated patches); moreover, model retraining is necessary whenever a new tool is to be added, 
which limits its flexibility and practicability. 

\begin{figure}[!tp]
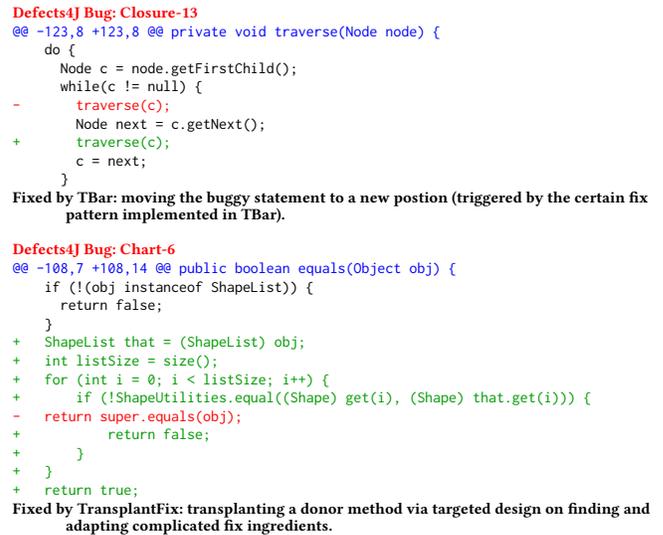

    \centering
    \vspace{0.2cm}
\begin{lstlisting}[language=diff,linewidth={\linewidth},frame=tb,basicstyle=\scriptsize\ttfamily]
(@{\color{red}\bf Defects4J Bug: Closure-13}@)
@@ -123,8 +123,8 @@ private void traverse(Node node) {
    do {
      Node c = node.getFirstChild();
      while(c != null) {
-       traverse(c);
        Node next = c.getNext();
+       traverse(c);
        c = next;
      }
(@{\bf Fixed by TBar: moving the buggy statement to a new postion (triggered by the certain fix pattern implemented in TBar).}@)

(@{\color{red}\bf Defects4J Bug: Chart-6}@)
@@ -108,7 +108,14 @@ public boolean equals(Object obj) {
    if (!(obj instanceof ShapeList)) { 
      return false;
    }
+   ShapeList that = (ShapeList) obj;
+   int listSize = size();
+   for (int i = 0; i < listSize; i++) {
+       if (!ShapeUtilities.equal((Shape) get(i), (Shape) that.get(i))) {
-   return super.equals(obj);
+           return false;
+       }
+   }
+   return true;
(@{\bf Fixed by TransplantFix: transplanting a donor method via targeted design on finding and adapting complicated fix ingredients.}@)
\end{lstlisting}
    \vspace{-0.3cm}
    \caption{Examples of two bugs fixed by APR tools with different bug-fixing strategies.}
    \label{fig:egs} 
\end{figure}

In this paper, we propose a novel ensemble strategy for APR that is motivated by the following hypothesis:
different APR tools achieve differing performance in bug fixing because they have different repair {\em preferences} (i.e., a feature set of bugs that an APR tool can fix).
Our hypothesis is formulated based on our examination of a  number of bug-fixing examples, two of which are shown in 
Figure~\ref{fig:egs}.
For example, Closure-13 is fixed by moving the buggy statement to a new position implemented in TBar \cite{issta19_TBar} with a certain fixing pattern, whereas
Chart-6, a bug that requires multi-line patches, is fixed by TransplantFix \cite{ase2022_transplantFix} via target design on finding and adapting complicated fix ingredients.
The preference that an APR tool has in fixing bugs is driven in part by the {\em repair pattern(s)} that the tool explicitly or implicitly employs as well as its {\em repair history}.


Given the above discussion, we propose \toolname (\underline{\textbf{P}}reference\underline{\textbf{-}}based \underline{\textbf{E}}nsemble \underline{\textbf{P}}rogram \underline{\textbf{R}}epair), a new ensemble strategy that 
leverages the repair preferences of each tool in the ensemble to improve repairability in practice.
\toolname is novel in the following respects:

{\bf No model training.} \toolname is the first {\em non-learning-based} ensemble method for APR. While E-APR casts the task as a multi-label {\em classification} task that involves identifying the subset of tools in the ensemble of tools for fixing a given bug, \toolname casts the task as a {\em ranking} task that ranks the tools in the ensemble based on how likely a tool can correctly fix a given bug. Moreover, unlike E-APR, which requires training a classification model in a supervised manner, \toolname does not require any model training. Specifically, \toolname ranks the tools in the ensemble by independently scoring each tool based on how likely it can fix the bug using several sources of information in a heuristic manner. 

{\bf New knowledge sources.} As mentioned before, \toolname employs two sources of information that encodes a tool's repair preferences, namely repair patterns and repair history. To our knowledge, the use of repair patterns has not been explored in existing ensemble methods for APR. Note that repair patterns encode a significant amount of human knowledge of the types of bugs a tool is adept at fixing and therefore they are likely to be more useful for scoring and ranking tools than any other program-independent and dependent features that one can possibly come up with. In fact, we believe that augmenting the feature set currently employed by E-APR with our repair patterns will likely boost its performance.

{\bf Dynamic updating.} As soon as a tool is used to fix a bug, \toolname receives {\em immediate} feedback on whether it successfully fixes the bug by having its repair history updated. In other words, the repair history of a tool is updated in a dynamic fashion based on all of the bugs that it has been applied to so far. Hence, \toolname has the ability to exploit information that it acquires in real time. This dynamic updating mechanism is one of the key strengths of \toolname that distinguishes it from existing ensemble methods for APR.

Another key strength of \toolname is its flexibility. One may argue that the need to manually identify repair patterns whenever a new non-learning-based tool is to be added to \toolname makes our approach undesirable or even impractical. It turns out that \toolname is flexible enough that one can add a new non-learning-based tool to it {\em without} identifying any repair patterns. In other words, the manual identification step is a {\em recommended} rather than {\em compulsory} step. To see the reason, recall that \toolname operates by scoring each tool w.r.t.\ a given bug using two sources of knowledge, repair patterns and repair history. If one source of information is absent (in this case the repair patterns), \toolname can simply  rely on the other source of information for scoring. In other words, the use of repair patterns can only improve ranking results, but \toolname can operate even without the patterns. 

In fact, \toolname is even more flexible than what we just described. In the extreme case, a new tool can be added to \toolname even when both sources of information (i.e., the patterns and the history) are absent. In this scenario, the new tool will receive a score of 0 at the beginning, but over time, the dynamic updating procedure will update its repair history\footnote{Given a bug, even a tool with \textit{repair patterns} and a \textit{repair history} may get a score of 0 if its \textit{repair patterns} and \textit{repair history} do not match the bug. Besides,  a tool may even get a score lower than 0 if its \textit{repair history} has many records of failing in fixing the current type of bug (since we will use the history of failures as a penalty). Therefore, a tool that has neither \textit{repair patterns} nor \textit{repair history} may be ranked higher than those with repair patterns and/or repair history.} In other words, over time, \toolname will be able to accumulate enough knowledge about the repair preferences of the new tool via updating its repair history even if we know nothing about it at the time of incorporation.

In sum, our work makes the following contributions. First,
    we propose the first non-learning-based ensemble strategy \toolname for assembling APR tools that is highly flexible.
%
    Second, we manually collect 13 repair patterns of APR tools, retrieve 4 types of bug features, and construct a mapping between repair patterns and bug features. Besides, we generate a categorized performance history between 21 concrete APR tools and the feature set of bugs that they can fix, which can be reused in other ensemble program repair frameworks. 
    Finally, we design specific evaluation metrics to measure the effectiveness of ensemble program repair strategies and conduct comprehensive experiments to evaluate \toolname. 

    Experimental results show that \toolname achieves better results than existing strategies. Two of the most significant empirical findings are that (1) when given the same amount of labeled data (which \toolname uses to initialize the repair histories of the tools and E-APR uses for model training), \toolname demonstrates that it is more effective at exploiting the labeled data by achieving considerably better results than E-APR; and (2) even when \toolname operates {\em witihout} using any repair patterns, it still outperforms E-APR, suggesting the robustness of \toolname.


\section{Background and Related Work}

\begin{figure*}
    \centering
    \setlength{\abovecaptionskip}{0cm}
    \includegraphics[width=\linewidth]{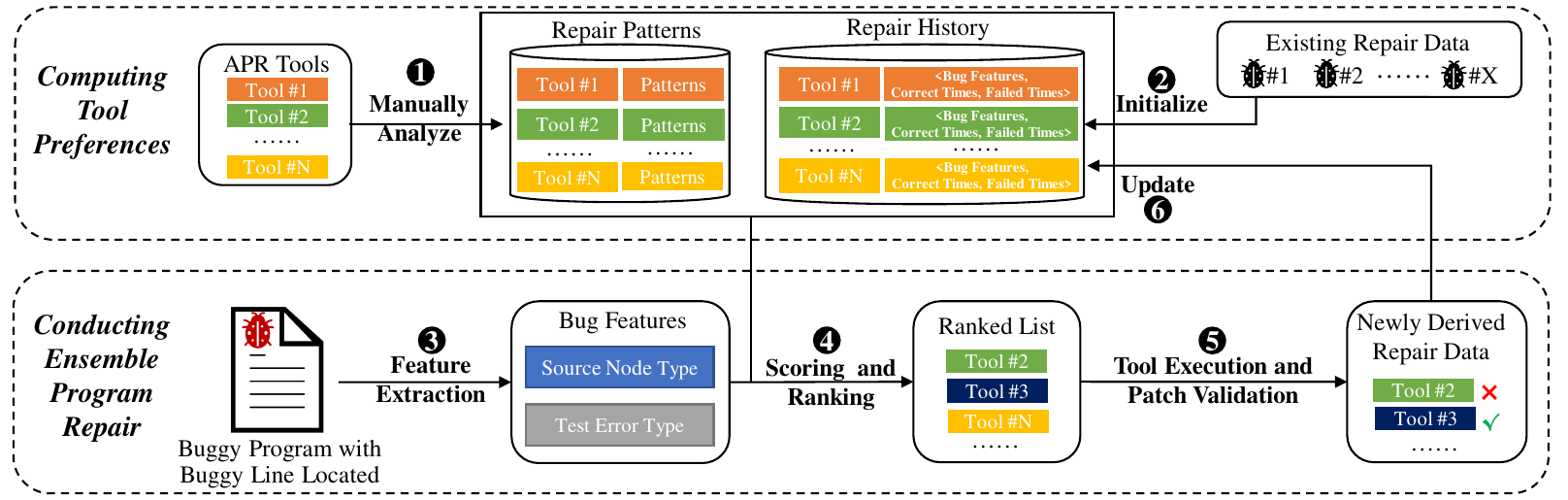}
    \caption{The overall procedure of \toolname.}
    \label{fig:EPRP_framework}
\end{figure*}

This section introduces the research background
of this work.

\subsection{Different Types of APR Tools}
Existing APR tools 
can be categorized into four types: 

{\bf Heuristic-based approaches} rely on manually defined heuristic rules to generate patches by iterating over a search space of syntactic program modifications, of which experimental results reviewed that they normally target on fixing general bugs \cite{icse09_genprog,tse12_genprog,icse14_RSRepair,issta16_astor,tse20_arja,issta18_simfix,ase2022_transplantFix}. However, they suffer from low efficiency due to the large search space and the limited effectiveness caused by the large number of plausible patches~\cite{survey19_APR}. 

{\bf Template-based approaches} generate patches based on a batch of pre-defined fix patterns, acting at explicit and direct modes \cite{icse13_PAR,issta16_astor,ase17_elixir,fse17_Genesis,saner17_npefix,icse18_sketchfix,icse18_CapGen,icst19_kPAR,saner19_avatar,icst19_kPAR,issta19_TBar,ese20_fixminer} : (1) checking whether the buggy statement satisfies the prepositive conditions of fix patterns, and (2) continuously generating code changes based on patterns until a valid patch is generated or the fixing behavior is terminated. Obviously, the repairability of template-based tools relies on the diversity of repair templates. 

{\bf Constraint-based approaches} use semantic constraints to limit the search space of patches \cite{tse17_nopol,icse16_Dynamoth,SSBSE18_Cardumen,fse17_S3} . Generally, such approaches first infer repair constraints from the buggy program or the test suite and use an SMT solver (e.g., Z3 \cite{TACAS08_Z3}) or other strategies to solve the constraints.
However, the symbolic execution and constraint solver can explode the space of generating constraints and patch candidates when fixing complex bugs.

{\bf Learning-based approaches} aim to train APR systems using historical bug-fixing data that can be sourced from code repositories.
For example, 
DeepRepair \cite{saner19_DeepRepair} relies on deep learning to sort repair ingredients via code similarities. 
Latest learning-based approaches \cite{tosem19_tufano, tse19_sequencer,issta20_coconut, ase20_Edits, tse2020_codit, icse21_cure, icse22_rewardrepair} employ neural machine translation (NMT) \cite{iclr15_NMT} models to perform bug-fixing framework as a sequence-to-sequence translation task.
Such methods rely on a large amount of bug-fix data and need to address the overfitting issue in the training process.  

\subsection{Empirical Studies on APR Tools}
Various empirical studies on APR tools have been conducted from different aspects to boost the development of automated program repair. 
Qi et al.~\cite{qi2015analysis} analyzed the correctness and plausibility of patches generated by APR tools.
Smith et al. \cite{esme19_eff_emp2} looked into the overfitting problem of patches generated by APR tools. 
Motwani et al. \cite{esme18_eff_emp3} investigated to what extent hard and important bugs can be fixed by APR tools.
Durieux et al. \cite{FSE19_empAPR} empirically studied the generalizability of 11 APR tools with five benchmarks and all possible repair attempts.
Liu et al. \cite{icse20_efficiency} explored the bug-fixing efficiency of 16 APR tools.
These studies demonstrated that different APR tools present varying repairability on different bugs and their specific characteristics of fixing bugs.

\subsection{Related Work}
As various APR tools are proposed, researchers have begun exploring advanced assembling methodologies to boost automated program repair, which can summarized into two categories: 

{\bf Exploiting multiple models.}
CoCoNut \cite{issta20_coconut} and CURE \cite{icse21_cure} are two learning-based techniques that 
train a neural APR model multiple times, each time with a different set of parameters. This results in multiple APR models. Each of these results is then used to independently generate patches. For example, 
if 10 models are trained and each one generates 300 patches, all 3000 patches will be validated. 
Hence, while these techniques generate an ensemble of models, strictly speaking they are {\em not} ensemble strategies.

{\bf Ensemble methods.} 
E-APR \cite{ese21_EAPR} makes an early exploration of reusing existing tools via an ensemble strategy. It first identifies significant program-independent and dependent features by analyzing footprints of repair results of existing APR tools. Then, it predicts the effectiveness of APR tools via machine learning algorithms according to the metrics of nine features identified from 146 features. However, a major limitation is that every time a new APR tool needs to be added to the ensemble, the model must be re-trained. In contrast, {\bf P-APR} does not require any model training.

\section{\toolname}
\label{sec:approach}

Figure \ref{fig:EPRP_framework} shows the overall procedure of \toolname, which consists of two parts: (1) \textbf{\textit{Computing Tool Preferences}} (Section 3.1), and (2) \textbf{\textit{Conducting Ensemble Program Repair}} (Section 3.2). 

\eat{
To compute tool preferences, we first manually analyze the target APR tools to collect their preferred operations in repairing bugs (i.e., \textbf{Step 1}). These patterns can be ultimately mapped to bug features through their corresponding pre-requirements on bugs. Next, if there are historically repaired bugs of target APR tools, we construct the repair history of tools categorized by bug features (i.e., \textbf{Step 2}). To fix a given bug, we extract its features (i.e., \textbf{Step 3}), and utilize the features and the tool preferences to score and rank target tools (i.e., \textbf{Step 4}). After sequentially executing tools and validating patches (i.e., \textbf{Step 5}), features of the newly fixed bugs will be used to update the Repair History (i.e., \textbf{Step 6}).
}

\subsection{Computing Tool Preferences}
\label{sec:prepare}

To capture the preference of an APR tool, we compute a tool's preferences in an offline manner (see the first part of Figure \ref{fig:EPRP_framework}) through two steps, as described below.




\begin{table*}[t]
\setlength{\abovecaptionskip}{0cm}
  \centering
  \caption{Example of repair patterns collected from existing APR systems.}
  \resizebox{\linewidth}{!}{
  \begin{threeparttable}

    \begin{tabular}{lp{11.375em}p{18.25em}p{25.915em}}
    \toprule
    \textbf{No.} & \textbf{Pattern Name} & \textbf{Pre-requirement} & \multicolumn{1}{l}{\textbf{Implemented APR systems}} \\
    \midrule
    \multirow{2}[2]{*}{P1} & \multirow{2}[2]{*}{Insert Cast Checker} & A buggy statement that contains  & heuristic-based: HDRepair, SimFix, CapGen \\
          & \multicolumn{1}{l}{} & at least one unchecked cast expression & template-based: AVATAR, Genesis, kPAR, SketchFix,  TBar, SOFix \\
    \midrule
    \multirow{3}[2]{*}{P2} & \multirow{3}[2]{*}{Insert Null Pointer Checker} & A buggy statement if, in this statement, a field & heuristic-based: HDRepair, SimFix, CapGen \\
          & \multicolumn{1}{l}{} & or  an expression (of non-primitive data type)  & tempate-based: AVATAR, Genesis, Elixir, FixMiner, NPEfix, SOFix \\
          & \multicolumn{1}{l}{} & is accessed without a null pointer check &                           kPAR, TBar \\
    \midrule
    \multirow{2}[2]{*}{P3} & \multirow{2}[2]{*}{Insert Range Checker} & Inserting a range checker for the access of an  & \multirow{2}[2]{*}{template-based: AVATAR, Elixir, kPAR, SketchFix, TBar, SOFix} \\
          & \multicolumn{1}{l}{} & array or collection if it is unchecked & \multicolumn{1}{l}{} \\
    \midrule
    P4    & Throw Exception & The failed test type is throwing an exception & heuristic-based: ACS \\
    \bottomrule
    \end{tabular}%
        \begin{tablenotes}
        \footnotesize
        \item[*] Due to the space limitations, we only list 4 patterns in this table. Concrete information of all 13 collected patterns can be found in the supplementary. 
    
      \end{tablenotes}
   \end{threeparttable}

    }

  \label{tab:pattern}%
\end{table*}%

\textbf{Step 1: Repair Pattern Collection via Manual Analysis.} 
We define a tool's repair action (e.g., Inserting Range Checker) for fixing a certain type of target bug as a Repair Pattern. Each repair pattern has corresponding pre-requirements on bugs (e.g., "If an array or collection is accessed without being checked" for Inserting Range Checker), which describes the characteristics of a bug that would trigger and be fixed by the repair pattern/action. An APR tool may have multiple repair patterns and one repair pattern may be shared by different APR tools. We summarize the repair patterns for existing APR tools by going through their methodologies and implementations. For example, for template-based tools, the repair patterns can be inferred directly from their implemented patterns. For other types of tools, a repair pattern can be searched by checking if the tool implements certain repair actions that must be triggered by conditions related to the input bugs. Note that there are no repair patterns defined in learning-based APR tools. 

Specifically, we collect all repair patterns by investigating the 42 APR tools listed in \cite{living_review} and \cite{issta19_TBar}. In total, we derive 13 repair patterns that are implemented in 19 non-learning-based APR tools. Table \ref{tab:pattern} presents the four example patterns, describing their name, pre-requirements and implemented APR systems.  

Next, we complete the judgement logic of whether an input bug satisfies the conditions of a certain repair pattern.
Motivated by a previous template-based APR tool TBar \cite{issta19_TBar}, we manually analyze the pre-requirements of the collected patterns and design four types of bug features (BF1-4) that could cover the automatic judgment logic of all patterns, as shown in Table \ref{tab:bugfeature}. For example, to check if a given bug satisfies \textit{P4 Throw Exception}, \toolname would examine whether BF4 (the type of the test error) of the bug is an \textit{Exception Thrown} error.

\textbf{Step 2: Repair History Initialization using Existing Repair Data.} 
Among the four features (BF1-4) defined in the previous step, BF1 (the node type of the buggy statement) and BF4 (the type of test error) can be regarded as program-independent since they can be extracted from any buggy program. Thus, we reuse BF1 and BF4 to initialize the repair history of a tool.
Concretely, we store a tuple \textit{<tool, bug\_feature, failed\_times, correct\_times>} for each tool, where \textit{bug\_feature} can be BF1 or BF4. 
When configuring an APR tool into \toolname, the existing repair history of the tool can be loaded to initialize the repair history. For example, APR tools are usually empirically evaluated in some bug-fix benchmarks before publication. Thus, before deploying \toolname in practice, those existing repair results can be utilized to enhance the performance of \toolname. Such design in \toolname ensures the generalization of \toolname on integrating any kinds of APR tools with existing repair results.

\subsection{Conducting Ensemble Program Repair}
\label{sec:conduct}
Given the Repair Patterns and the Repair History of the APR tools involved, \toolname can be used to repair a given bug. The input of \toolname is the buggy class file along with its suspicious faulty lines located by fault localization techniques \cite{fl2016survey}.

\begin{table}[t]
  \centering\
   \setlength{\abovecaptionskip}{0cm}
  \caption{Bug features needed for matching different patterns}
  \resizebox{\linewidth}{!}{
  \begin{threeparttable}            
    \begin{tabular}{lll}
    \toprule
    \textbf{No.} & \textbf{Bug Feature } & \textbf{Pattern*} \\
    \midrule
    BF1     & Node type of the buggy statement & P6, 11, 12  \\
    BF2     & Child node types within the buggy statement & P3, 5, 7-9 \\
    BF3     & BF1 \& BF2 & P1, 10 \\
    BF4     & Type of the test error & P4 \\
    \bottomrule
    \end{tabular}%
         \begin{tablenotes}
        \footnotesize
        \item[*] Corresponding descriptions of each pattern can be found in the supplementary.
        We ignore P2 since its pre-requirement is too general that almost every statement can match the pattern.
    
      \end{tablenotes}
 \end{threeparttable}

    }
    
  \label{tab:bugfeature}%
\end{table}%

\IncMargin{1em}
\begin{algorithm}
     \caption{Calculate preference scores for APR tools}
     \label{alg:preference-score}
     \SetAlgoNoLine
     \KwInput{The faulty line IDs, $allFaultyLineIds$}
     \KwInput{The faulty class file, $buggyFile$}
     \KwInput{The bonus coefficient of pattern match, $EM_\alpha$}
     \KwOutput{The preference scores of all tools, $preferScores$ }
     $preferScores \leftarrow \varnothing$ \;
     $EM_\alpha \leftarrow 0.5$ \;
     \For {$lineId \in allFaultyLines$}{
     $bugFeatures \leftarrow ExtractFeature$ ($lineId$, $buggyFile$)\;
         \For {tool $\in$ availableTools}{
         $finalScore \leftarrow 0$\;
         $historyScore \leftarrow 0$\;
         \For {feature $\in$ bugFeatures}{
         $historyScore \leftarrow $historyScore + $CalculateHistoryScore(tool$, $feature$)\;
         }

         \eIf{$PatternMatch(tool, bugFeatures)$}{
         $finalScore \leftarrow historyScore * (1 + EM_\alpha )$\;
         }{ $finalScore \leftarrow historyScore$\;}
           $preferScores.set(tool,finalScore)$\;  
         }
     }
\end{algorithm}

\textbf{Step 3: Feature Extraction.} 
To automatically determine whether the input bug satisfies the repair patterns and repair history of APR tools, we need to first extract features of the given bug. 
Recall that we define 4 features for bugs in Table \ref{tab:bugfeature}.
Bug features 1, 2, and 3 are properties of code elements within the buggy statement. We use spoon \cite{spoon} to parse the buggy code and extract those features. 
Bug feature 4 corresponds to the type of failed test error triggered by the bug, such as \textit{java.lang.IndexOutOfBoundsException}. Trivial test error, i.e., \textit{junit.framework.AssertionFailedError}, is disregarded since it is too general. If a bug produces multiple failed test cases, only the first non-trivial test error is considered. 

\eat{vince
其实这个motivation不是针对现有工作的不足之处而产生的。就是在introduction的第4段和figure 2，因为发现用了不同bug-fixing strategies的工具能针对性修复不同的bug，所以产生了用repair pattern来指导工具选择的想法
ACS: constraint-based, TBar: template-based, TransplantFix: heuristic-based
}

\textbf{Step 4: Tool Scoring and Ranking.} 
Algorithm \ref{alg:preference-score} presents the core step of \toolname, where a score representing the chance that the given bug can be correctly fixed is calculated for each APR tool. Concretely, the score is derived by matching the prepared tool preferences (i.e., including Repair Pattern and Repair History) with features of the given bug. To score each APR tool, we use the faulty code file and the faulty line IDs as inputs. 
The algorithm is compatible with bugs containing any number of hunks. For each faulty line, the bug features defined in Table \ref{tab:bugfeature} are first extracted (line 4). Then, for each tool in the available toolset, we calculate a $historyScore$ according to the existing repair history of the tool (line 8-10). 
Recall that \toolname stores the repair history of a tool with a tuple \textit{<tool, bug\_feature, fail\_times, correct\_times>}, so the $historyScore$ is calculated as: 
\begin{equation}
\begin{aligned}
& CalculateHistoryScore(tool,feature) = \\ &correct\_times_{tf}/(correct\_times_{tf}+fail\_times_{tf}) 
\end{aligned}
\end{equation}
where $t$ represents the tool and $f$ represents the feature. We index the repair history with BF1 and BF4. For example, if the node type of the buggy statement is \textit{CtInvocationImpl}, the corresponding preference score will be the fixed rate when the tool encounters bugs of such type. Then, \toolname will judge if the bug features match the preferred patterns of tools in the available set (line 14). If yes, the preference score of the tool will get a bonus (line 15).
At this step, we introduce a configurable coefficient $EM_\alpha$ to control the bonus degree. We use multiplication to combine the pattern and history preference scores.
The final preference score of each tool is the sum of the preference score of all faulty lines. The higher the score is, the more likely it is for the corresponding tool to fix the bug. All the tools are ranked in descending order of scores. 

\textbf{Step 5: Tool Execution and Patch Validation.} 
After the tools are ranked in descending order of scores for a bug, a human developer can use the tools sequentially to fix the bug. However, it is not necessary to use these tools sequentially. For instance, if there are $K$ available computing threads/human developers available to fix the bug with APR tools at the same time, the Top-$K$ tools could be adopted simultaneously. It is worth mentioning is that generating patches and checking if a patch is plausible is achieved by the tool automatically, but checking if a plausible patch is a correct one can only be achieved by a human developer. This is 
why we calculate two different costs for fixing a bug when evaluating \toolname. 

\textbf{Step 6: Preference Update.} 
As aforementioned, the history score is computed according to the repair history of the APR tools on bugs that have the same features as the given bug. Therefore, at the end of each repair procedure, \toolname updates the repair history of each tool according to their performance on the bug that they just handled if they are executed. In our implementation of \toolname, we maintain a repair result list of each APR tool in the available toolset, and each repair result is represented by <$FixStatus$,$BugFeatures$,$Tool$>, where $FixStatus$ denotes the repair status with three enum types (correct, overfit, fail) and $BugFeatures$ identifies the characteristics of the buggy program. Like in Repair History, we use BF1 and BF4 defined in Table \ref{tab:bugfeature} as the content of $BugFeatures$ in repair results. 

\subsection{Integrating Improved or New APR Tools}
\label{sec:generalizability}

To integrate an improved or new tool into \toolname, we need to perform the two steps below:%
\footnote{For more concrete instructions of integrating improved or new APR tools into \toolname, as well as using \toolname, please refer to the tool's repository: https://github.com/kwz219/P-EPR-Artefact}: 

\vspace{-2mm}
\paragraph{(1) Updating Repair History.} This means updating the Repair History table shown in Figure~\ref{fig:EPRP_framework} for the target APR tool with buggy programs that have been repaired by the target tool either successfully or unsuccessfully. Users are only required to provide \textcircled{1} the buggy class file, \textcircled{2} the suspicious line locations, and \textcircled{3} the test error type (if available). \toolname first transforms the given buggy programs into bug features (shown in Table~\ref{tab:bugfeature}) and then updates the corresponding tuple \textit{<tool, bug\_feature, fail\_times, correct\_times>} for the tool. Any buggy program adopted by any APR tool as repair history can be integrated into our Repair History table since our bug feature BF1 is program-independent, i.e., any buggy program has a value for BF1. In other words, our \toolname can be generalized to any improved or new APR tool with any kind of repair history regardless of whether it has Repair Patterns. 

\vspace{-2mm}
\paragraph{(2) Updating Repair Patterns.} 
This step is needed only when the improved/new APR tool to be added is non-learning-based. Given the improved/new tool, we need to identify the set of repair patterns associated with it. If all of the repair patterns it is associated with are among the 13 patterns that currently exist in \toolname, then nothing needs to be done. otherwise, for each new pattern, we need to add it to the pattern repository and update the mapping of patterns to bug features (see the current mapping in Table~2). 

\section{Evaluation Setup}

\subsection{Research Questions}
\label{sec:rqs}
We aim at answering the following three research questions for evaluating \toolname. 

\textbf{[RQ1. Performance] What is the overall performance of \toolname compared with the other ensemble strategies?} Concretely, we conduct different ensemble strategies to select APR tools for each bug in the Defects4J v1.2 dataset. To thoroughly evaluate \toolname, we consider a maximum set of 21 tools.
Note that we do not execute each tool to derive the progress due to the unaffordable costs. Instead, we rely on all published patches of different tools for each bug in Defects4J v1.2.

\textbf{[RQ2. Ablation Study]  To what extent does each component of \toolname contribute to its overall performance?} We seek to gain insights into \toolname by understanding the impacts of its components on the performance, such as the Test error type and the coefficient $EM_{\alpha}$, via ablation experiments. 

\textbf{[RQ3. Practicality] To what extent can \toolname save computational costs in practice compared with adopting every single tool?} 'in practice' means executing the selected tool by \toolname to calculate the computational cost (e.g., time for generating patches, time for verifying patches, and computer memory, etc.) instead of performing simulations using existing patches. Considering that the Defects4J dataset has been used by almost all APR tools and that the performance of \toolname on Defects4J is evaluated in RQ1, we use another dataset, Bears, to verify the performance of \toolname in practice (where the repair history of the APR tools on Defects4J are used for initializing their Repair History in \toolname).

\subsection{Tool Selection and Data Collection}\label{sec:data_collection}
\begin{table}[t]
  \centering
  \setlength{\abovecaptionskip}{0cm}
  \caption{Correct/overfit patches generated by the 21 APR systems on 395 bugs from Defects4J v1.2  }
\resizebox{0.8\linewidth}{!}{
    \begin{tabular}{rlccr}
\cmidrule{2-5}          & \textbf{System} & \textbf{\# Correct} & \textbf{\# Overfit} & \multicolumn{1}{l}{\textbf{Source}} \\
\cmidrule{2-5}    \multicolumn{1}{c}{\hspace*{-4pt}\multirow{9}{*}{\rotatebox[origin=c]{90}{heuristic-based}}} & jGenProg & 6     & 10    & \multicolumn{1}{c}{\multirow{8}[1]{*}{\cite{icse20_efficiency}}} \\
          & GenProg-A & 8     & 21    &  \\
          & RSRepair-A & 9     & 25    &  \\
          & ARJA  & 11    & 25    &  \\
          & SimFix & 29    & 21    &  \\
          & jKali & 2     & 6     &  \\
          & Kali-A & 5     & 37    &  \\
          & jMutRepair & 5     & 6     &  \\
          & TransplantFix & 36    & 33    & \multicolumn{1}{c}{\cite{ase2022_transplantFix}} \\
\cmidrule{2-5}    \multicolumn{1}{c}{\hspace*{-4pt}\multirow{4}{*}{\rotatebox[origin=c]{90}{constraint}}} & Nopol & 2     & 7     & \multicolumn{1}{c}{\multirow{4}[1]{*}{\cite{icse20_efficiency}}} \\
          & ACS   & 16    & 5     &  \\
          & Cardumen & 2     & 14    &  \\
          & DynaMoth & 3     & 10    &  \\
\cmidrule{2-5}    \multicolumn{1}{c}{\hspace*{-4pt}\multirow{4}{*}{\rotatebox[origin=c]{90}{template}}} & kPAR  & 33    & 30    & \multicolumn{1}{c}{\multirow{4}[1]{*}{\cite{icse20_efficiency}}} \\
          & AVATAR & 30    & 20    &  \\
          & FixMiner & 34    & 29    &  \\
          & TBar  & 54    & 30    &  \\
\cmidrule{2-5}    \multicolumn{1}{c}{\hspace*{-4pt}\multirow{4}{*}{\rotatebox[origin=c]{90}{learning}}} & SequenceR & 27    & 24    & \multicolumn{1}{c}{\multirow{4}[1]{*}{\cite{ase2022_npr4j} }} \\
          & CodeBERT-ft & 29    & 28    &  \\
          & RewardRepair & 43    & 22    &  \\
          & Recoder & 56    & 22    &  \\
\cmidrule{2-5}          & Total & 122   & 121(58) &  \\
\cmidrule{2-5}    \end{tabular}%
}
  \label{tab:datacollection}%
\end{table}%

Since \toolname is compatible with any kind of APR tool, we select 
a variety of APR tools. However, empirically executing a large number of APR tools and validating generated patches is prohibitively expensive. So, we choose to evaluate the performance of \toolname through a simulated experiment with published repair results of APR tools, instead of actually running APR tools. Simulation means that we directly get the repair results of APR tools on each bug, skipping the tool execution and patch validation process. To reduce the biases brought by the simulated experiment, we select APR tools for the simulated experiment according to the following criteria:

\textbf{C1: The fault localization setting of each APR tool should be the same.} Since \toolname extracts features of faulty code lines, different fault localization results can impact the calculated score of the same bug. However, it is challenging to maintain the same fault localization setting when considering Normal Fault Localization (NFL). This is because NFL settings of existing APR tools vary significantly on the FL tool and considered fault locations. For example, SketchFix \cite{icse18_sketchfix} considers only the top 50 most suspicious statements in the ranked list, while ELIXIR \cite{ase17_elixir} considers up to the top 200 suspicious locations. 
Therefore, to minimize biases in our experiment, we only considered tools evaluated within the Restricted Fault Localization (RFL) scenario \cite{icse20_efficiency}, where the accurate faulty line of the buggy program is provided. 

\textbf{C2: The patch generation setting of each APR tool should be the same.}
To satisfy this criterion, we opt to refer to empirical studies on APR tools, which typically use the same settings across the studied tools, instead of collecting repair results from individual APR publications. We first obtain the repair results of 16 test-suite-based APR tools from a relevant empirical study \cite{icse20_efficiency}. Additionally, we include four state-of-the-art learning-based tools in our evaluation by re-running them on Defects4J with NPR4J \cite{ase2022_npr4j}, a framework tool that supports running these tools. 

\begin{table*}[t]
  \centering
  \setlength{\abovecaptionskip}{0cm}
  \caption{\toolname's performance on 395 bugs from Defects4J considering 21 APR tools. For each bug, the Top-$K$ APR systems ranked by \toolname are selected to repair it. 
  "Opt" and "All" denote Optimal Selection and Invoking All Tools strategies respectively.}
  \resizebox{\linewidth}{!}{
  \begin{threeparttable}
          \begin{tabular}{lccccccccc|cc}
    \toprule
          & \textbf{Top-1} & \textbf{Top-2} & \textbf{Top-3} & \textbf{Top-4} & \textbf{Top-5} & \textbf{Top-6} & \textbf{Top-7} & \textbf{Top-8} & \textbf{Top-9} & \textbf{Opt} & \textbf{All} \\
    \midrule
    \textbf{\# of correctly/plausibly fixed bugs} & 54/89 & 68/100 & 78/113 & 87/129 & 86/133 & 95/138 & 101/146 & 108/157 & 109/160 & 122/180 & 122/180 \\
    \textbf{\# of plausible patches} & 108   & 159   & 219   & 278   & 326   & 389   & 453   & 478   & 510   & 180   & 859 \\
    \textbf{Tool Invocation Times (TISP)} & 1010 (33\%) & 1461 (39\%) & 1925 (41\%) & 2330 (44\%) & 2701 (38\%) & 3102 (41\%) & 3488 (41\%) & 3906 (42\%) & 4282 (38\%) & 395 (95\%) & 8295 \\
    \textbf{Human Valdation Times (HVSP)} & 98 (20\%) & 117 (26\%) & 148 (27\%) & 175 (27\%) & 191 (22\%) & 214 (24\%) & 229 (25\%) & 257 (24\%) & 271 (21\%) & 180 (54\%) & 393* \\
    \bottomrule
    \end{tabular}%
    
     \begin{tablenotes}
        \footnotesize
        \item[*] When invoking all tools, the plausible patches are ordered in the same way as the tools were initially added to the toolset. 
        \item[**] Due to space limitations, we only present results from top-1 to top-9. Results of larger $K$ values are listed in the GitHub repo. 
      \end{tablenotes}
    \end{threeparttable}
    }
  \label{tab:performance}%
\end{table*}%

Given these  criteria, we collect the repair results of 21 APR systems on Defects4J v1.2. 
Those systems cover 4 types of APR tools: 9 are \textit{heuristic-based} (jGenProg \cite{issta16_astor}, GenProg-A \cite{tse20_arja}, RSRepair-A \cite{tse20_arja}, ARJA \cite{tse20_arja}, SimFix \cite{issta18_simfix}, jKali \cite{issta16_astor}, Kali-A \cite{tse20_arja}, jMutRepair \cite{issta16_astor}, TransplantFix \cite{ase2022_transplantFix}), 4 are \textit{constraint-based} (Nopol \cite{tse17_nopol}, ACS \cite{icse17_ACS}, Cardumen \cite{SSBSE18_Cardumen}, DynaMoth \cite{icse16_Dynamoth}), 4 are \textit{template-based} (kPAR \cite{icst19_kPAR}, AVATAR \cite{saner19_avatar}, FixMiner \cite{ese20_fixminer}, TBar \cite{issta19_TBar}) and 4 are \textit{learning-based} (SequenceR \cite{tse19_sequencer}, CodeBERT-ft \cite{msr21_CodeBERT}, RewardRepair \cite{icse22_rewardrepair}, Recoder \cite{fse21_recoder}). Among the 21 systems, we re-run the four learning-based tools (for execution and validating settings, ref to Section \ref{sec:setting}) since they do not provide required data (i.e., both correct and overfit patches generated by the tool) and use the published patches of other systems. In total, the 21 o tools correctly/plausibly fix 122/180 bugs from Defects4J. 9 of them have repair patterns (SimFix, jMutRepair, Nopol, ACS, Dynamoth, kPAR, AVATAR, FixMiner, TBar).

\subsection{Metrics}
\label{sec:metrics}
First, to estimate the repairability and costs when deploying \toolname on a set of APR tools to fix bugs, we use the following metrics:

    \textbf{(1) The number of correctly/plausibly fixed bugs.} A plausible patch can pass all test cases, but it may not be correct. A correct patch can pass all test cases and human validation.
    
    \textbf{(2) Tool Invocation Times (TIT).} It measures the machine resource costs when invoking a set of APR tools. For simplicity, we define one tool invocation as whether the APR tool should be invoked when a bug is given. 
    
    \textbf{(3) Human Validation Times (HVT).} It measures the human labor costs of checking plausible patches. For a bug, a tool selection strategy may generate more than one plausible patch. We define HVT as the number of manual checks needed to find a correct patch. If no correct patches are generated, the HVT is equal to the number of generated plausible patches. 

Second, to quantify the cost savings obtained by employing \toolname, we design two novel metrics:

    \textbf{(4) Tool Invocation Saving Percentage (TISP).} It measures how many tool invocation times can be saved when using a tool selection strategy compared with invoking all APR tools. TISP is calculated as:
    \begin{equation}
TISP = R_{Strategy}/R_{EnsAll} - TIT_{Strategy}/TIT_{EnsAll}
\end{equation}
    where $R_{Strategy}$ and $R_{EnsAll}$ represent the numbers of correctly fixed bugs of using a strategy and of invoking all available tools respectively, and $TIT_{Strategy}$ and $TIT_{EnsAll}$ denote the tool invocation times of using a strategy and of invoking all available tools respectively.
    
    \textbf{(5) Human Validation Saving Percentage (HVSP).} It measures how many manual checks can be saved when using a tool selection strategy compared with invoking all APR tools. HVSP is calculated as:
    \begin{equation}
HVSP = R_{Strategy}/R_{EnsAll} - HVT_{Strategy}/HVT_{EnsAll}
\end{equation}
    where $R_{Strategy}$ and $R_{EnsAll}$ are the same as those in Equation~2, and $HVT_{Strategy}$ and $HVT_{EnsAll}$ denote the human validation times of using a strategy and of invoking all available tools respectively.

\subsection{Baseline Systems}
We compare \toolname with several baselines: 

\textbf{(1) E-APR \cite{ese21_EAPR}}. Since E-APR's source codes are not published, we replicate it. Specifically, we implement E-APR with Random Forest Classifier since it achieves the best performance among the four algorithms described in E-APR's paper.

\textbf{(2) E-APR (enhanced)}. The original E-APR is trained and tested on only 10 APR tools. We consstruct an enhanced version of E-APR by re-training it using the same 21 tools that \toolname uses using
a Random Forest Classifier following the settings of E-APR.

\textbf{(3) Random Selection}. This strategy randomly selects $K$ tools sequentially for each bug as the top-ranked tools.

\textbf{(4) Invoking All Tools}. This strategy 
simply invokes all available tools to fix each bug.

\textbf{(5) Optimal Selection (Ground Truth)}. This is an ideal 
strategy that 
makes the optimal choice of APR tools, thus providing a rough upper bound on \toolname's performance. It prioritizes the tools as follows: \textcircled{1} can produce correct patches, \textcircled{2} can produce plausible patches, and \textcircled{3} cannot produce plausible patches. 

\subsection{Tool Execution and Validation Settings}\label{sec:setting}
Our experiments involve practical execution of four APR tools (SequenceR \cite{tse19_sequencer}, CodeBERT-ft \cite{msr21_CodeBERT}, Recoder \cite{fse21_recoder} and RewardRepair \cite{icse22_rewardrepair} ) on Defects4J v1.2 \cite{ISSTA14_defects4J} for the data collection in previous section,  and Bears \cite{SANER19_bears} for an empirical experiment in RQ3. We use the same tool execution and patch validation settings. For each bug, each tool generates 300 candidate patches, with a timeout of 2 hours for validating them. For patch correctness assessment, two of the authors manually validate the first test-adequate patch with a timeout of 10 minutes for every bug, adhering to the assessment criteria established by prior research \cite{icse20_efficiency}. A patch is deemed correct only if both reviewers agree on its accuracy. All model execution and patch validating experiments are performed on a machine equipped with an AMD Ryzen 9 5950X 16-Core Processor and two NVIDIA GeForce RTX 3090 Ti GPUs.

\section{Evaluation Results}


\subsection{RQ1. Performance of \toolname}

\textit{Method:} We design three experiments. First, we execute \toolname on all of the 395 bugs from Defects4J using the 21 APR tools. We use all data in Defects4J as test data, effectively assuming that no bugs are used for initializing the repair history. Second, we compare \toolname with the original E-APR \cite{ese21_EAPR} (which is trained and evaluated on only 10 of the 21 APR tools) by only using the 10 APR tools used by the original E-APR. The third one involves comparing \toolname with the enhanced E-APR, which is re-trained on all of the 21 APR tools.
In the second and third experiments, we also compare with the Random selection strategy.
Besides, in these experiments, we assume that the input order of bugs is random (random seed = 1) and $EM_\alpha$ (see \textbf{Step 4} in Section \ref{sec:conduct}) is 0.5 (the default setting)~\footnote{Since the dynamic update module updates the repair history of APR tools in \toolname according to the input bug and the repair history affects the performance of \toolname in real time, the input order of test bugs will affect the evaluation results of \toolname. Besides, \textit{the different combinations of APR tools} may also affect the evaluation of the overall performance of \toolname. So, we conduct additional experiments by setting them with different values to demonstrate the feasibility of \toolname. See these experimental results and discussions in Section 2 in the supplementary file.}. 

\textit{Results and discussion:} Next, we describe and discuss the results of the aforementioned expriements.

\begin{figure}[t]
\centering
\setlength{\abovecaptionskip}{0cm}
\subfigure{
\setlength{\abovecaptionskip}{0cm}
\includegraphics[width=0.9\linewidth]{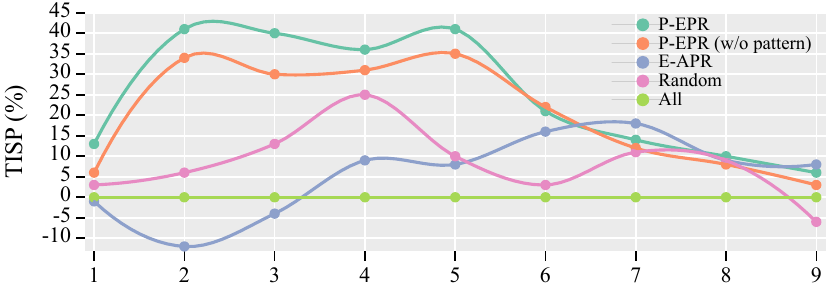}
}
\quad
\subfigure{
\setlength{\abovecaptionskip}{0cm}
\includegraphics[width=0.9\linewidth]{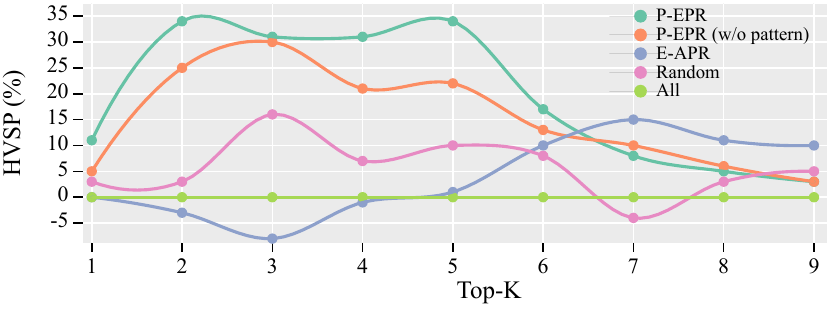}

}
\caption{ Comparison of \toolname and other strategies in terms of TISP on Defects4J employing only the 10 APR tools originally used by E-APR. "All" denotes invoking all APR tools. The HVSP of the optimal strategy is 57\%, and the TISP of the optimal strategy is 90\%.}
\label{fig:compare_EAPR}
\end{figure}

\begin{figure}[t]
\centering
\setlength{\abovecaptionskip}{0cm}
\subfigure{
\setlength{\abovecaptionskip}{0cm}
\includegraphics[width=\linewidth]{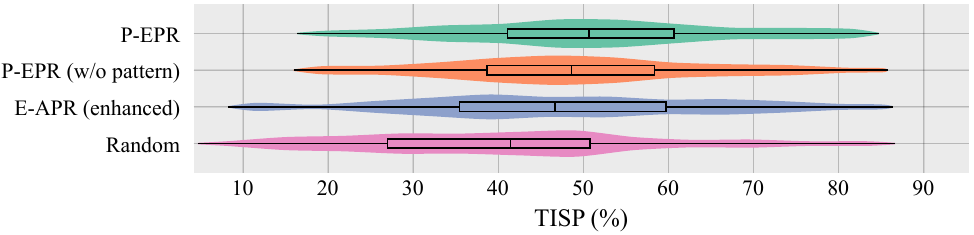}
}
\quad
\subfigure{
\setlength{\abovecaptionskip}{0cm}
\includegraphics[width=\linewidth]{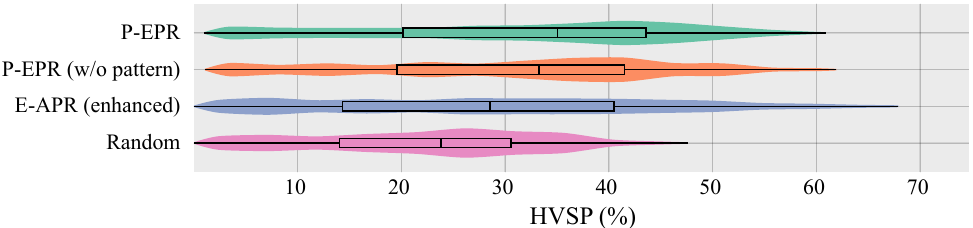}
}
\caption{ Distributions of TISP and HVSP of \toolname and other strategies on the different train-test split of Defects4J projects and different \textit{K} (ranges from 1 to 20) considering 21 tools. }
\label{fig:distribution}
\end{figure}

\eat{
Table \ref{tab:performance} represents the comprehensive performance (including all metrics mentioned in Section \ref{sec:metrics}) of \toolname considering 21 APR tools on Defects4J. Figure \ref{fig:compare_EAPR} display the comparison (in terms of TISP and HVSP) of \toolname, original E-APR, and random selection strategy on 10 APR tools. Figure \ref{fig:distribution} displays the distributions of TISP and HVSP of \toolname, enhanced E-APR, and random selection strategy considering 400 situations (i.e., there are 20 train-test splits situations of selecting three projects from Defects4J's six projects for initializing repair history in \toolname, and for each bug, we could select Top-$K$ tools to fix it, $K$ ranges from 1 to 20, so, for ecah strategy, 400 TISPs and 400 HVSPs should be calculated and plotted in Figure 5). 
}

\textbf{Comparison with Invoking All Tools and the Optimal Strategy.} Table \ref{tab:performance} presents the repair results (the number of correctly/plausibly fixed bugs), execution costs (model invocation times and human validation times), and costs saving a percentage of \toolname (TISP and HVSP) when selecting Top-1 to Top-9 to repair bugs from Defects4J when all 21 tools are used, as well as the corresponding metrics of Invoking All Tools and Optimal Selection (i.e., Ground Truth) strategies.
Compared with simply executing all available APR tools, \toolname can significantly save costs on tool execution and human validation while reaching comparable repairability. 
For example, when \textit{K} is 9, \toolname achieves a 90\% repairability compared with executing all tools (109 correctly fixed bugs for \toolname vs.\ 122 for all tools), with a 50\% reduction of execution costs in terms of the model invocation times (4282 vs.\ 8295) and a 30\% reduction of human validation costs (271 vs.\ 393). As \textit{K} decreases, \toolname also achieves remarkable performance. When \textit{K} is 1, \toolname correctly fixes 54 bugs, which is very close to the best APR tool in our experiment (Recoder \cite{fse21_recoder} fixes 56 bugs). When \textit{K} is larger than 2, \toolname outperforms any one of the 21 tools significantly. 
Remembering that we have a large-scale set of 21 APR tools, the results prove that \toolname has the ability to correctly rank APR tools among a variety of tools. 
Compared with the optimal strategy, \toolname achieve 35\% - 44\% of the optimal strategy on TISP, and 37\% to 50\% on HVSP. It indicates that the performance of the current \toolname strategy still has a large room for optimization.

\textbf{Comparison with the original E-APR.} 
Figure \ref{fig:compare_EAPR} expresses the performance of \toolname, the original E-APR, and the random selection strategy in terms of TISP and HVSP when only the 10 APR tools employed by the original E-APR are involved.
As can be seen, \toolname achieves significantly higher performance than E-APR and random selection. In terms of TISP, \toolname always performs better than the random selection. Compared with E-APR, \toolname has a significant improvement when \textit{K} ranges from 1 to 5. 
In some cases E-APR performs worse than just invoking all APR tools. When \textit{K} is 2, \toolname has the highest improvement on TISP than E-APR (40\% vs -15\%). In terms of HVSP, \toolname performs better in most cases than other strategies. Besides, when \textit{K} is between 2 and 5, \toolname achieves the highest TISP and HVSP, while E-APR's performance improves as \textit{K} becomes larger. This suggests that a higher level ensemble strategy of \toolname and E-APR may yield even better performance. 


\textbf{Comparison with enhanced E-APR.}
In the previous comparisons, we used all of Defects4J as test data, leaving no training data that can be used to initialize a tool's repair history. In this comparison, we perform comparisons where we do partition Defect4J into a training set and a test set so that we can initialize the repair history of each tool in \toolname using the training data. More specifically, we show in Figure \ref{fig:distribution} the distributions of TISP and HVSP of \toolname computed based on 400 situations (i.e., there are 20 train-test splits situations of selecting three projects from Defects4J's six projects, and for each bug, we could select Top-$K$ tools to fix it, $K$ ranges from 1 to 20; so, for each strategy, 400 TISPs and 400 HVSPs are calculated and plotted in Figure 5).
We similarly plot the curves for two other baselines, enhanced E-APR and the random selection strategy 
Recall that enhanced E-APR is the version of E-APR that is being retrained on all the 21 APR tools used in \toolname. For a fairer comparison with \toolname, E-APR is re-trained using not only the training data used in the original E-APR but also the training portion of Defect4J in each of the 400 situations mentioned above. As shown in Figure \ref{fig:distribution}, \toolname still outperforms enhanced E-APR.
In terms of both TISP and HVSP, \toolname has a larger minimum value, Q1 (first quartile), Q2 (second quartile), Q3 (third quartile) than enhanced E-APR and random selection. Also, \toolname has a shorter IRQ (Inter Quartile Range) than enhanced E-APR, which means that generally, \toolname has better and more stable performance than the two strategies. However, we also observe that \toolname has a lower maximum value than enhanced E-APR (especially TISP). 

\begin{figure*}[t!]
\centering
\setlength{\abovecaptionskip}{0cm}
\subfigure[Impact of Pattern, Dynamic Update, Repair History Initialization and Test Error Type on TISP.]{
\includegraphics[width=3.98cm]{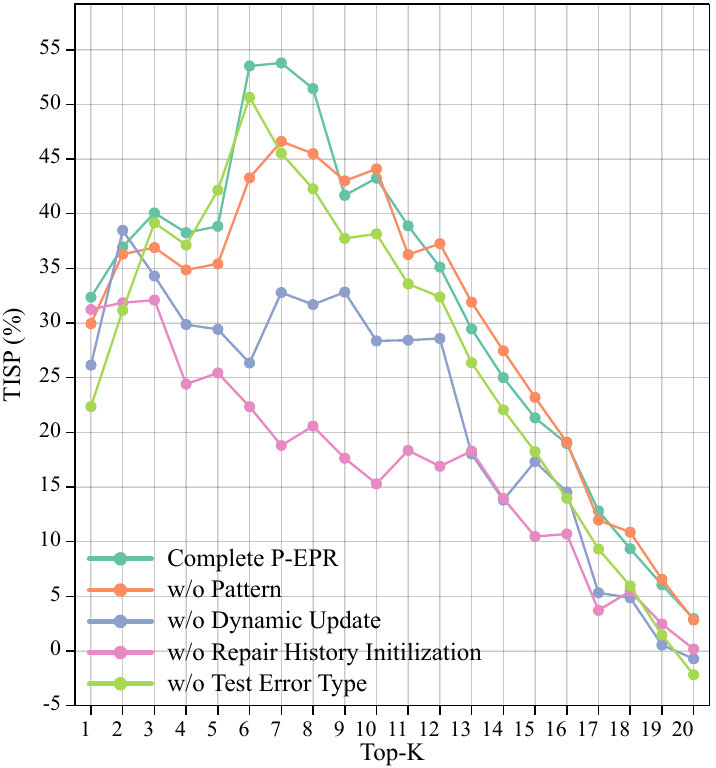}
}
\quad
\subfigure[Impact of different $EM_\alpha$ on TISP.]{
\includegraphics[width=3.98cm]{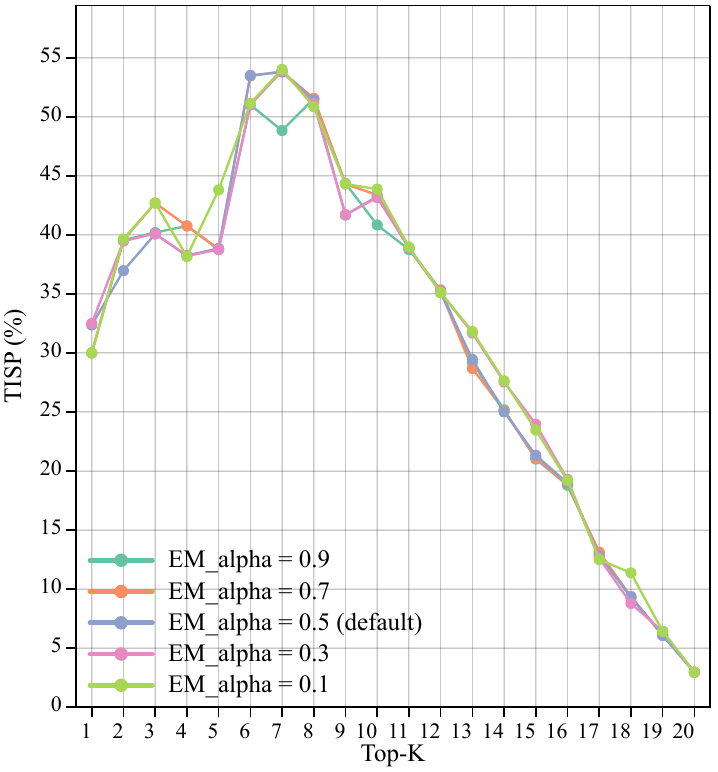}
}
\quad
\subfigure[Impact of Pattern, Dynamic Update, Repair History Initialization and Test Error Type on HVSP.]{
\includegraphics[width=3.98cm]{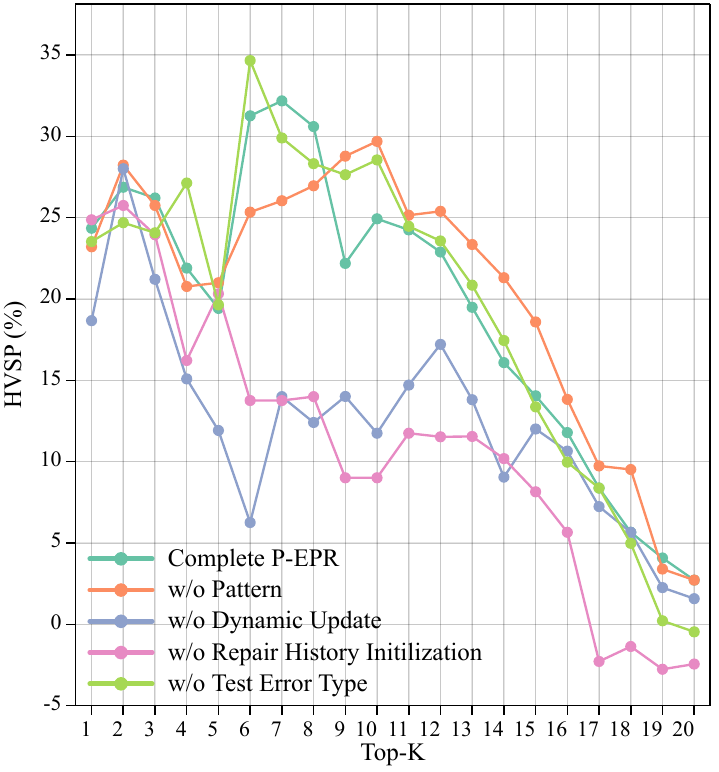}
}
\quad
\subfigure[Impact of different $EM_\alpha$ on HVSP.]{
\includegraphics[width=3.98cm]{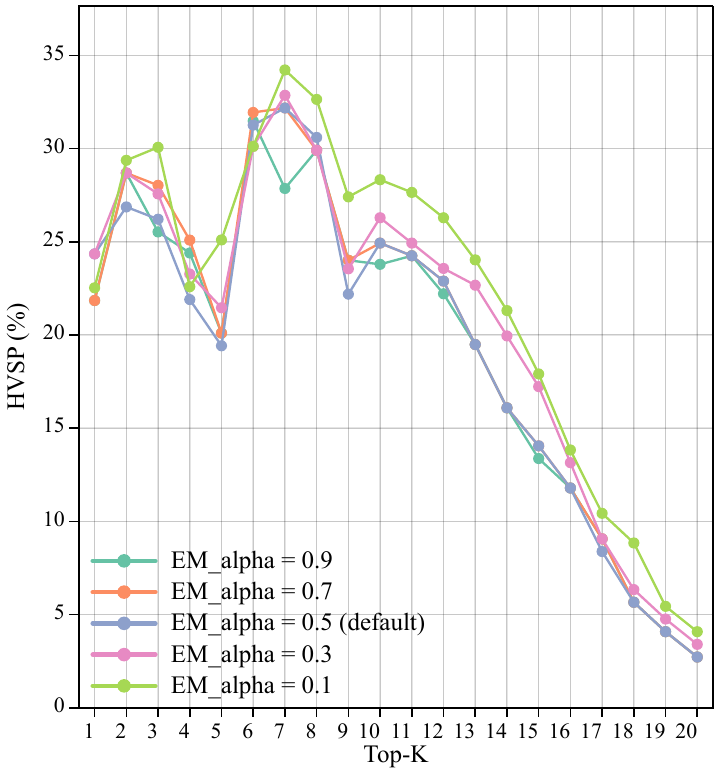}
}
\caption{Ablation experiments illustrating the impact of different components of \toolname when used with 21 APR tools.
}
\label{fig:ablation}
\end{figure*}

\begin{table*}[t]
  \centering
  \setlength{\abovecaptionskip}{0cm}
  \caption{Performance and Computational Costs of \toolname on the Bears benchmark. Machine Validation Time represents time costs on executing test cases. Human Validation Times represents time costs on manually checking plausible patches.  }
  
  \begin{threeparttable}
    \resizebox{0.85\linewidth}{!}{
    \begin{tabular}{l|cccc|ccc}
    \toprule
    \multirow{2}[4]{*}{\textbf{Metrics}} & \multicolumn{4}{c|}{\textbf{Single System}} & \multicolumn{3}{c}{\textbf{Selection Strategy}} \\
\cmidrule{2-8}          & \textbf{Recoder} & \textbf{CodeBERT-ft} & \textbf{RewardRepair} & \textbf{SequenceR} & \textbf{P-EPR (top-1)} & \textbf{Optimal*} & \textbf{All*} \\
    \midrule
    \# of correct/plausible patches & 10/21 & 12/25 & 12/21 & 14/28 & 16/28 & 22/37 & 22/37 \\
    Precision & 48\%  & 48\%  & 57\%  & 50\%  & 57\%  & 59\%  & / \\
    Inference Time (min) & 11    & 2     & 4     & 4     & 7     & 6     & 28 \\
    Machine Validation Time (hour) & 57    & 54    & 58    & 52    & 52    & 40    & 221 \\
    Human Validation Time (min) & 41    & 54    & 36    & 62    & 56    & 84    & 118 \\
    TISP  & 20\%  & 29\%  & 29\%  & 38\%  & 48\%  & 75\%  & / \\
    HVSP  & 5\%   & 7\%   & 15\%  & 10\%  & 20\%  & 29\%  & / \\
    GPU Memory Usage (GB) & 19.1  & 3.84  & 7.32  & 8.17  & 7.83  & /     & / \\
    \bottomrule
    \end{tabular}%
    }
    \begin{tablenotes}
        \footnotesize
        \item[*] We use the information produced when individually executing each program repair tool to estimate the performance and costs of invoking all tools. For example, we record the machine and manual validation time when individually executing each system and use them to estimate corresponding metrics when performing the two strategies. For the strategy that invokes all tools, we assume tools are sequentially executed.     
    \end{tablenotes}
   
  \end{threeparttable}
  \label{tab:emp_result}%
\end{table*}%



\subsection{RQ2. Contributions of Components}
\textit{Method:} We investigate the impacts of each components by comparing \toolname with following variants: (1) \toolname without Pattern, (2) \toolname without Dynamic Update, (3) \toolname without Repair History Initialization, (4) \toolname without Test Error Type and (5) \toolname with different  $EM_\alpha$ (0.1, 0.3, 0.5, 0.7, 0.9). For comparison, we select the Math project (contains 106 bugs) from the Defects4J project as the evaluation set and use the other five projects (contains 289 bugs) as the repair history for initialization (except variant (3)). We use the 21 APR tools and calculate their TISP and HVSP.   
It should be mentioned that there is no need to conduct any ablation experiment on the he Buggy element type feature (i.e., BF1) because all bugs have this feature and it is obvious that it plays an important role in ranking APR tools for bugs. Considering that while adding a new APR tool into \toolname, collecting its repair patterns may be a little bit harder than collecting its repair history, we also conduct ablation experiments on Repair Patterns while comparing \toolname with the other strategies to investigate whether the performance of \toolname will be lower than the other strategies if we want to save the cost of collecting repair patterns of the new APR tools. 

\textit{Results and discussion:} Figures \ref{fig:ablation} (a) and (c) show the comparison between variants (1) - (4) in terms of TISP and HVSP, respectively. As can be seen, Repair History Initialization and Dynamic Update have significantly higher impacts than Repair Patterns and Test Error Type in \toolname. Without repair history initialization, \toolname has a up-to-50\% performance degradation both in terms of TISP and HVSP. \toolname without Dynamic Update suffers a similar performance degradation. The two components contributes to \toolname via categorizing and updating history, which indicates that \toolname benefits significantly from the repair data. It is recommended that a good practice for users to utilize \toolname is to provide data for initialization and keep updating with newly derived repair results. 

In addition to Figures~\ref{fig:ablation} (a) and (c), which show the relative small impact of Repair Patterns on \toolname, Figures~\ref{fig:compare_EAPR} and ~\ref{fig:distribution} 
show that even without Repair Patterns, \toolname still outperforms the other ensemble strategies. 
This implies that \toolname can still help us select appropriate APR tools for bugs if we want to save the cost of collecting repair patterns of APR tools. However, these figures also show that the complete \toolname outperforms the variant without Pattern in most cases when \textit{K} ranges from 1 to 9. This implies that we should encourage the users to analyze patterns when integrating new tools into \toolname for a even better ensemble performance.

 
In Figure~\ref{fig:ablation} (a) and (c), we see that the variant without Test Error Type seems has relatively smaller performance difference with complete \toolname comparing against other components. It suggests that \toolname can be generalized in more scenarios (e.g., the bug is identified by a bug report but not a test case failure). 

Figure \ref{fig:ablation} (b) and (d) compare the impacts of different $EM_\alpha$ values (variant 5) in terms of TISP and HVSP. As can be seen, the value of $EM_alpha$ has little impact on the performance of \toolname, which means users can pay less attention to setting this parameter.  

\subsection{RQ3. Practicality of \toolname}

\textit{Method:} 
We collect 83 single-hunk bugs from another dataset, i.e., Bears~\cite{SANER19_bears} for investigating the performance of \toolname while practically executing integrated APR tools. We configure \toolname with four learning-based program repair systems (i.e., SeqeunceR \cite{tse19_sequencer}, CodeBERT-ft \cite{msr21_CodeBERT}, RewardRepair \cite{icse22_rewardrepair}, and Recoder \cite{fse21_recoder})~\footnote{
Recall that the reason of conducting simulation experiments to evaluate the performance of \toolname on 21 APR tools is that it is prohibitively expensive for us to empirically executing a large number of APR tools and validating generated patches. So, in the evaluation of practicality, we try our best to include APR tools into the experiments considering the computational and human resources that we can afford. 
} 
and initialize \toolname with their existing repair results on Defects4J collected beforehand. We set $EM_\alpha$ to 0.5 by default. During inference, we select the top-1 system to repair every bug and record the Inference Time, Machine Validation Time, Human Validation Time, and GPU Memory Usage. In cases where multiple systems have the same top-1 score, we always choose the system with the least GPU usage (i.e., CodeBert-ft < SequenceR < RewardRepair < Recoder).

\textit{Results and discussion:} Table \ref{tab:emp_result} illustrates the performance and costs associated with executing each tool individually, as well as deploying \toolname to select the top-ranked tool for execution. As depicted, \toolname achieves the highest levels of repairability, successfully fixing 16 bugs, and demonstrates superior precision (57\%) compared to the other four tools and the optimal selection strategy. In comparison to the strategy of executing all tools, \toolname achieves a repairability of 76\% (16 out of 21 bugs) while maintaining a higher precision, at a significantly reduced cost (25\% of inference time and 24\% of machine validation time). On human validation times, \toolname also has a lower cost at 67\% and 47\%, compared with the optimal strategy and invoking all tools. 
This case serves as a concrete example of \toolname's effectiveness in bug fixing, highlighting its feasibility and practical application.

\section{Discussion}

{\bf Practicality.} One may be concerned that the design of patterns involves a large amount of manual work and is prone to human errors. While this is a valid concern, there are a few things to keep in mind. First, patterns are applicable for non-learning-based tools. Given that the majority of work on APR these days are learning-based, we expect that our patterns need to be updated on an occasional basis. Second, 
these repair patterns can be reused in other ensemble program repair frameworks. For example, it is conceivable that these patterns can be encoded as features that can be used to augment the feature sets used in existing learning-based ensemble methods for APR such as E-APR, and given the rich amount of human knowledge these patterns encode, they are likely to be useful for other ensemble program repair frameworks as well. Third, \toolname is flexible enough that it can be deployed without any repair patterns (with the caveat that performance may suffer as a result). While errors could be introduced in the derivation of patterns, one could employ a second person to verify the correctness of the patterns.

Some may argue that we should instead go for a learning-based ensemble method for APR in which we train a model using program-independent and dependent features that can generalize the results to future improvements. This is exactly what is done in E-APR. While a systematic analysis of learning- and non-learning-based ensemble approaches to APR is beyond the scope of this paper, there are a few things that we should keep in mind. First, while we acknowledge the importance in developing program-independent and dependent features, these features by no means render our repair patterns useless. Specifically, these features and the repair patterns encode different kinds of knowledge.
It is not even clear whether learning a model using only program-independent and dependent features will ever perform as well as one that uses patterns as features given the rich amount of human knowledge encoded in the patterns. Nevertheless, our results indicate that E-APR underperforms \toolname when given the same amount of labeled data. Second, when incorporating a new tool to a learning-based ensemble method for APR, one needs to provide a possibly large amount of labeled data so that the model can learn how to classify/rank the new tool against the existing tools. In other words, if one believes that the amount of manual effort that goes into the identification of patterns makes \toolname less practical, then the amount of manual effort that goes into providing labeled data may similarly make a learning-based approach impractical. Finally, while at first glance it seems that with a learning-based approach we can focus on developing program-independent and dependent features that can generalize the results to future improvements, the non-generalizable part of a learning-based approach is hidden in the manually labeled, tool-specific training data. In other words, for any ensemble approach to APR, there has to be a non-generalizable component that is specific to the new tool to be added, either in the form of labeled data (for a learning-based approach) or as explicitly stated repair patterns (as in \toolname).

{\bf Performance metrics.} 
TIT and HVT are not entirely representing the related costs, thus the newly derived TISP and HVSP can not precisely measure the practical performance of \toolname. However, the real tool execution and labor costs are also hard to be fairly and precisely measured even when executing the APR tools due to various factors such as machine performance and reviewer proficiency. Since no standard metrics are previously proposed for measuring the APR tool selection strategy, our proposed metrics contributes by providing an easy-to-compute though still imperfect quantified measurement for comparing different tool selection strategies.

{\bf Implications for practitioners.}
Our research has several implications. Firstly, our analysis demonstrates that the repair preference difference of current APR tools can be distinguished by simple bug features and test error types. Results of our experiments in RQ3 show practical cost savings through the deployment of \toolname for selecting APR tools. It is important to note that the current version of \toolname only optimizes the repair probability of different bugs. To further improve the tool selection strategy, researchers can consider incorporating additional execution information, such as GPU memory usage and test feedback during validation. This avenue holds promise for designing a better approach.
Secondly, \toolname provides a simple and extensive way to leverage existing APR tools for enhanced repair performance, at a lower cost on tool execution and patch validation compared with invoking all tools. For configuring and extending new APR tools to \toolname, users only need to provide the APR tool's repair history and implemented pattern type (if available). This strategy can also benefit some scenarios where tools must be run locally.
For instance, recent studies have explored bug repairs using large language models such as CodeX \cite{CodeX-repair} and ChatGPT \cite{ChatGPT-repair}. However, these methods may raise security concerns since the model holders only provide a remote API, which means users must post their codes to the remote host. Instead, our approach enables users to achieve comparable performance by locally executing existing tools.

\section{Threats to Validity}
Threats to external validity include the evaluated dataset used in our experiment, i.e. Defects4J. We only evaluate \toolname considering APR tools of Java on 395 bugs from Defects4J. However, repairing Java programs is the most popular research scene for the APR community and Defects4J is the most popular dataset. Besides, we evaluate \toolname on a variety of APR tool combinations (up to 21 tools), which could alleviate the threats to some extent. 

That we choose to perform a simulated experiment instead of executing APR tools could be a threat to internal validity. To reduce the threat, we collect repair results of existing APR tools following strict criteria to avoid biases brought by fault localization and patch generation settings. For learning-based tools that do not publish complete correct/plausible patches for the bugs used in our experiments, we re-run them following their experimental configurations in their papers or source code to collect the complete correct/plausible patches they generate for these bugs. Another threat is that the input orders of bugs could impact the performance of \toolname. We mitigate it by conducting rich experiments under different input orders of bugs in RQ2.

\section{Conclusion}
We presented a practical approach, referred to as \toolname, for selecting the most suitable automated program repair tools for a given software bug. \toolname is designed as a flexible and tunable framework that can interface with any type and quantity of APR tools to align with users' preferences. We evaluated its effectiveness and generalizability using a variety of tool combinations (up to 21 APR tools) on the Defects4J dataset. Additionally, we proposed two novel metrics that measure the extent to which a model selection strategy can reduce tool invocation and human validation costs compared with invoking all tools.
Our study demonstrated the potential for selecting optimal APR tools for distinct bugs, thus offering a novel and practical avenue for future research.


\section{Acknowledgement}
This research / project is supported by the National Key Research and Development Program of China (2022YFF0711404), National Natural Science Foundation of China (62172214), Natural Science Foundation of Jiangsu Province, China (BK20201250, BK20210279), CCF-Huawei Populus Grove Fund, NSF award 2034508, and the European Research Council (ERC) under the European Union's Horizon 2020 research and innovation program (grant agreement No. 949014). Any opinions, findings and conclusions or recommendations expressed in this material are those of the author(s). We also thank the reviewers for their helpful comments. Chuanyi Li and Jidong Ge are the corresponding authors.

\bibliographystyle{ACM-Reference-Format}
\bibliography{sample-base}

\appendix

\end{document}


\title{Supplementary}

 \maketitle

\paragraph{\textbf{Algorithm for Tool Scoring and Ranking}}

\IncMargin{1em}
\begin{algorithm}
    \caption{Calculate preference scores for APR tools}
    \label{alg:preference-score}
    \SetAlgoNoLine
    \KwInput{The faulty line IDs, $allFaultyLineIds$}
    \KwInput{The faulty class file, $buggyFile$}
    \KwInput{The bonus coefficient of pattern match, $EM_\alpha$}
    \KwOutput{The preference scores of all tools, $preferScores$ }
    $preferScores \leftarrow \varnothing$ \;
    $EM_\alpha \leftarrow 0.5$ \;
    \For {$lineId \in allFaultyLines$}{
        $finalScore \leftarrow 0$\;
        $historyScore \leftarrow 0$\;
        \tcp{extract bug features}
    $bugFeatures \leftarrow ExtractFeature$ ($lineId$, $buggyFile$)\;
        \For {tool $\in$ availableTools}{
        \tcp{calculate history preference scores}
        \For {feature $\in$ bugFeatures}{
        $historyScore.add($ $CalculateHistoryScore(tool$, $feature$) )\;
        }

       \tcp{If match a pattern, bonus score }
        \eIf{$PatternMatch(tool, bugFeatures)$}{
        $finalScore \leftarrow historyScore * (1 + EM_\alpha )$\;
        }{ $finalScore \leftarrow historyScore$\;}
        \tcp{sum scores of all faulty lines}
          $preferScores.update(tool,finalScore)$\;  
        }
    }
\end{algorithm}

This is the core step of P-ERP, where a score representing the ability/possibility of correctly fixing the input bug should be calculated for each APR tool. Concretely, the score is derived by matching the prepared tool preferences (i.e., including Repair Pattern and Repair History) with features of the input bug. Algorithm \ref{alg:preference-score} shows the complete procedure of scoring each APR tool with the faulty code file and the faulty line IDs as inputs. 
The algorithm is compatible with bugs containing any number of hunks. For each faulty line, the bug features defined in Table \ref{tab:bugfeature} are firstly extracted (line 6). Then, for each tool in the available toolset, we first calculate the  $historyScore$ according to the existing repair history of the tool (lines 9-11). 
Recall that \toolname stores the repair history of a tool with a tuple \textit{<tool, bug\_feature, fail\_times, correct\_times>}, the $historyScore$ is calculated as (ref to line 11):
\begin{equation}
\begin{aligned}
& CalculateHistoryScore(tool,feature) = \\ &correct\_times_{tf}/(correct\_times_{tf}+fail\_times_{tf}) 
\end{aligned}
\end{equation}
where $t$ represents the tool and $f$ represents the feature. We index the repair history with BF1 and BF4. For example, if the node type of the buggy statement is \textit{CtInvocationImpl}, the corresponding preference score will be the fixed rate when the tool encounters such type of bugs (line 10). Then, \toolname will judge if the bug features match the preferred patterns of tools in the available set (line 13). If so, the preference score of the tool will get a bonus (line 14).
At this step, we introduce a configurable coefficient $EM_\alpha$ to control the bonus degree. We use multiplication to combine the pattern and history preference scores.
The final preference score of each tool is the sum of the preference score of all faulty lines. The higher the score is, the higher the possibility of a tool to fix the bug. Eventually, all the tools are ranked in descending order of scores.

\appendix